\documentclass[apj]{emulateapj}

\shortauthors{Rogers et al.}

\usepackage{graphicx}
\usepackage{epstopdf}
\usepackage{epsfig}
\usepackage{natbib}
\usepackage{wasysym}
\begin{document}

\title{Three Possible Origins for the Gas Layer on GJ~1214b}

\author{L. A. Rogers\altaffilmark{1} and S. Seager\altaffilmark{1,}\altaffilmark{2}}

\altaffiltext{1}{Department of Physics, Massachusetts Institute of Technology, Cambridge, MA 02139, USA}
\altaffiltext{2}{Department of Earth, Atmospheric, and Planetary Sciences, Massachusetts Institute of Technology, Cambridge, MA 02139, USA}

\begin{abstract}

We present an analysis of the bulk composition of the MEarth transiting super Earth exoplanet GJ~1214b using planet interior structure models.  We consider three possible origins for the gas layer on GJ~1214b: direct accretion of gas from the protoplanetary nebula, sublimation of ices, and outgassing from rocky material. Armed only with measurements of the planet mass ($M_p=6.55\pm0.98~M_{\oplus}$), radius ($R_p=2.678\pm0.13~R_{\oplus}$), and stellar irradiation level, our main conclusion is that we cannot infer a unique composition. A diverse range of planet interiors fits the measured planet properties. Nonetheless, GJ~1214b's relatively low average density $(\rho_p=1870\pm400~\rm{kg\,m^{-3}})$ means that it almost certainly has a significant gas component. Our second major conclusion is that under most conditions we consider GJ~1214b would not have liquid water. Even if the outer envelope is predominantly sublimated water ice, the envelope will likely consist of a super-fluid layer sandwiched between vapor above and plasma (electrically conductive fluid) below at greater depths. In our models, a low intrinsic planet luminosity $\left(\lesssim2~\rm{TW}\right)$ is needed for a water envelope on GJ~1214b to pass through the liquid phase. 

\end{abstract}


\section{Introduction}

The  era of super Earths is upon us with the first two transiting sub-Neptune mass exoplanets recently discovered. The first such transiting planet, CoRoT-7b, is a $M=4.8\pm0.8~M_\oplus$ \citep{LegerEt2009A&A} and $R=1.68\pm0.09~R_\oplus$ \citep{QuelozEt2009A&A} hot relatively dense planet with an average density similar to Earth's.
More recently, the MEarth project \citep{IrwinEt2009IAUS} discovered transiting low-mass planet GJ~1214b \citep{CharbonneauEt2009Nature}. GJ~1214b has a mass of $M_p=6.55\pm0.98~M_{\oplus}$ and a radius of $R_p=2.678\pm0.13~R_{\oplus}$. It is in a $1.5803952\pm0.0000137$ day period around an $L_*=0.00328\pm0.00045~L_{\odot}$ M dwarf of mass $M_*=0.157\pm0.019~M_{\odot}$ and radius $R_*=0.2110\pm0.0097~R_{\odot}$.

GJ~1214b has a low enough density $(\rho_p=1870\pm400~\rm{kg\,m^{-3}})$ that it cannot be composed of rocky and iron material alone. The planet almost certainly contains a gas component. Even a planet of pure water ice is still too dense to match the observed mass and radius. At $6.55~M_{\oplus}$ a pure zero-temperature water ice planet would have a radius of $2.29~R_{\oplus}$ while an Earth-like composition would have a radius of about $1.64~R_{\oplus}$; these theoretical radii are 3 and $8\sigma$ lower than the value measured for GJ~1214b. While these simple arguments already reveal that GJ~1214b probably has a gaseous component, we are motivated to provide a more detailed analysis to quantify the range of possible planetary interior and gas layer compositions for GJ~1214b.

We use planet interior structure models to constrain the bulk composition of GJ~1214b. In this work, we focus on three possible sources for the GJ~1214b gas layer: direct accretion of gas from the protoplanetary nebula, sublimation of ices, and outgassing from rocky material. We examine end-member cases in which one of these three contributions dominates the gas layer. Based on GJ~1214b's mass and radius alone, we cannot infer a unique interior composition \citep[see, e.g.][]{ValenciaEt2007bApJ, AdamsEt2008ApJ, Zeng&Seager2008PASP}. Instead, there is a range of compositions that are consistent with the transit and radial velocity observations. Despite the inherent degeneracies plaguing the under-constrained problem of inferring GJ~1214b's composition from its mass and radius, we can nonetheless place interesting bounds on the gas envelope mass and draw insights into GJ~1214b's prospects for harboring liquid water.

In Section~\ref{sec:glo} we explore the connection between the primordial material comprising a planet and the sources of a planet's gas envelope. In Section~\ref{sec:model} we describe our model of low-mass planet interiors. In Section~\ref{sec:results} we present constraints on the composition of GJ~1214b in each of three distinct scenarios for the origin of its gas layer. Discussion and conclusions follow in Sections~\ref{sec:dis} and \ref{sec:con}.

\section{Connecting Gas Layer Origins and Planet Interiors}
\label{sec:glo}

There is a wide range of possible chemical compositions for  GJ~1214b's interior and gas layer. To motivate the discrete representative scenarios considered in this work, we look to the broad phases of materials that can contribute to a planet's bulk and to its gas layer.

\begin{figure}
\plotone{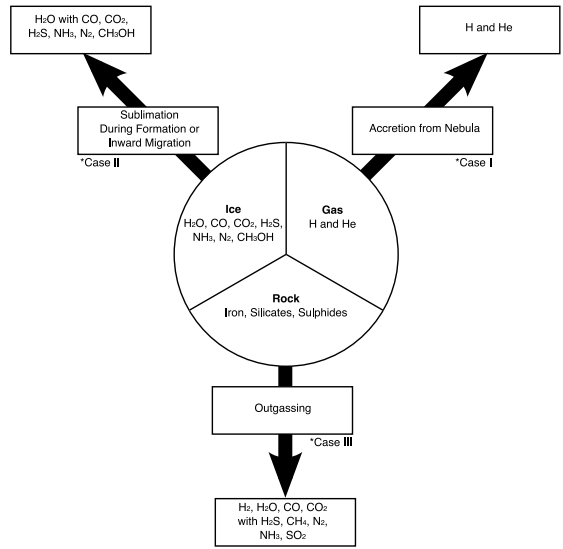}
\caption{Diagram detailing possible sources considered in this work for GJ~1214b's gas layer. The segments of the circle represent the three categories of primordial material that could have contributed to forming GJ~1214b: refractory materials, ice-forming material, and nebular gas. All three categories of primordial material can contribute to an eventual planetary gas layer. The gas formation processes we consider for GJ~1214b are indicated by the black arrows. Each arrow points to a box describing possible initial chemical compositions for the gas contributed by each source. The ice compositions were taken from \citet{MarboeufEt2008ApJ}, and the outgassed atmosphere compositions reflect the chemical equilibrium results of \citet{Schaefer&Fegley2009bAstroPh}. The gas layer composition will evolve over time under the influence of atmospheric escape. While primordial gas, ice, and rocks are all given equal fractions of the circle in this diagram, the primordial gas-ice-rock ratios of planets can vary over a wide range and affect the relative importance of the three gas layer sources shown in the diagram. In this work, we consider three end-member cases (labeled as cases I, II, and III) in which a single gas layer source dominated on GJ~1214b; the case associated with each gas formation process is indicated in the diagram.}
\label{fig:flowchart}
\end{figure}

GJ~1214b may have formed from a variety of primordial material in its protoplanetary disk including gas (predominantly hydrogen and helium); ice-forming material (water, carbon monoxide, carbon dioxide, methane, and ammonia); and rocks or refractory material (iron, silicates, and sulfides). All three classes of primordial planet-building material (nebular gas, ices, and rocks) could have contributed to the gas layer observed on GJ~1214b today (Figure~\ref{fig:flowchart}). Gas accreted directly from the nebula during planet formation, if retained, would contribute hydrogen and helium. Sublimation of ices (for example, due to the release of gravitational energy during initial planet formation, the increased stellar irradiation following inward planetary migration, or late delivery of ices by comets) would produce H$_2$O, CO, CO$_2$, CH$_4$, and NH$_3$ vapor. Finally, rocky material can release volatiles to the GJ~1214b gas layer via outgassing during formation \citep{ElkinsTanton&Seager2008aApJ, Schaefer&Fegley2009bAstroPh} and tectonic activity after formation \citep{KiteEt2009ApJ}. Irrespective the origin of GJ~1214b's gas layer (be it from accreted nebular gas, sublimated ices, or outgassed rocky material), the gas envelope's mass and composition will have evolved over time under the influence of atmospheric escape. 

In this work, we focus on direct accretion of nebular gas, sublimation of ices, and outgassing from rocky material as possible sources for the gas layer on GJ~1214b. There are, however, other atmosphere formation processes worth mentioning. Vaporization of rocky material can contribute to atmospheres surrounding highly irradiated super Earths like CoRoT-7b~\citep{Schaefer&Fegley2009ApJ}, but GJ~1214b is not hot enough for this process to occur. Even at temperatures too low to sublimate ices or vaporize oceans, volatiles stored as clathrate hydrates in icy material can be outgassed into a planet's atmosphere. Sputtering by the stellar wind and micrometeorites, photolysis, radiolysis, and chemical reactions between stellar wind ions and planet surfaces all contribute to tenuous atmospheres surrounding solar system bodies. The contributions of these gas sources are negligible, however, compared to the gas volume needed to account for GJ~1214b's transit depth.

\begin{figure}
\epsscale{1}
\plotone{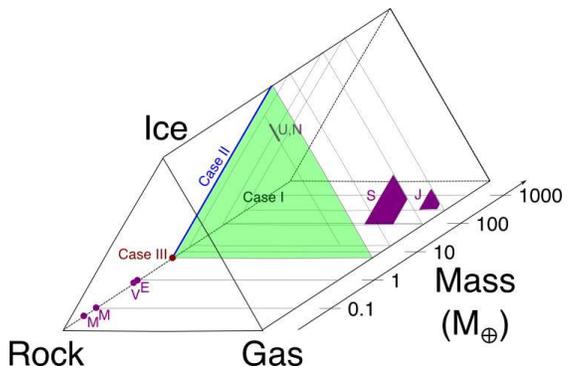}
\caption{Schematic diagram illustrating the range of possible planet  primordial bulk compositions. In this figure ``gas" refers to primordial H and He accreted from the nebula, ``ice" refers to ice-forming materials, and ``rock" refers to refractory materials (e.g., iron and silicates). Constraints on the current compositions of the solar system planets are plotted in purple (planets are denoted by their first initial). For GJ~1214b, we consider the full possible range of primordial gas, ice and rock relative abundances. In case I (green), GJ~1214b accreted and retained primordial gas, ices and refractory material. In case II (blue), GJ~1214b did not retain any primordial gas, incorporating only icy and rocky materials. Finally in case III (red), GJ~1214b formed from purely rocky material. Planets in cases II and III, which do not contain any primordial gas, may still harbor a gas layer produced by sublimated ices or by outgassing. This diagram was inspired by \citet{Chambers2010Book} and \citet{Stevenson2004PhT}.
}
\label{fig:prism}
\end{figure}

The actual bulk make-up of GJ~1214b is determined by its unknown formation, migration, and evolution history. For instance, if GJ~1214b initially formed beyond the snow line it would contain more icy material than if it formed closer to its star. The mass of nebular gas initially captured would depend upon the nascent GJ~1214b's accretion luminosity as well as the local conditions (density, temperature, opacity, and mean molecular weight) in the protoplanetary disk \citep{Rafikov2006ApJ}. In the solar system, there is a definite relationship between the relative abundances of rock-ice-gas and planet mass: small planets ($\leq1~M_{\oplus}$) are rocky, intermediate planets ($\sim 15-17~M_{\oplus}$) are icy, and larger planets are predominantly composed of H and He. Rough constraints on solar system planet compositions from \citet{Guillot2005AREPS} are plotted in Figure~\ref{fig:prism}. We do not attempt to tighten the constraints on GJ~1214b's interior by directly incorporating planet formation theories into our analysis. Instead we allow for the full range of primordial gas-ice-rock ratios and explore the constraints imposed on these ratios by the measured mass and radius.

We consider a series of scenarios that encompasses all  nebular gas-ice-rock mass fraction combinations  for the  primordial material making up GJ~1214b (Figure~\ref{fig:prism}). In each case, we assume that a single contributor (nebular gas, ice, and rock) dominated as the source for the gas envelope observed today on GJ~1214b. We are thus considering end-member scenarios within the continuum of possible gas envelope compositions. In case I, GJ~1214b managed to accrete and retain H and He gases from the nebula, and includes primordial gas, ice, and rock in its bulk make-up. In this scenario, we neglect any contributions to the gas envelope from the ices or rock, and take the current gas envelope on GJ~1214b to be composed of H and He. In case II, GJ~1214b has an interior formed from icy material and rock and did not retain any nebular gas (either having never accreted any in the first place or having lost any nebular gas that it once had). Here, we assume that vapors from the icy material dominate the gas layer (neglecting any contributions from the rocky material). Finally, in case III, GJ~1214b formed from purely rocky material and did not acquire any ices or gas from the protoplanetary disk.  In the absence of accreted ices and gas, the planetary gas envelope must originate by outgassing during formation or tectonic activity.

\section{Structure Model}
\label{sec:model}

Our model for the interior structure of low-mass exoplanets is described in detail in \citet{Rogers&Seager2010ApJ}. Here, we give a brief summary.
	
	We assume spherically symmetric and differentiated planets consisting of a core, mantle, ice layer, and gas envelope. The equation for the mass of a spherical shell and the equation describing hydrostatic equilibrium form a coupled set of differential equations for the radius $r\left(m\right)$ and pressure $P\left(m\right)$, viewed as functions of the interior mass $m$,
\begin{eqnarray}
\frac{dr}{dm}&=&\frac{1}{4\pi r^2\rho}\label{eqn:dr}\\
\frac{dP}{dm}&=&-\frac{Gm}{4\pi r^4},\label{eqn:dP}
\end{eqnarray}
\noindent where $\rho$ is the density and $G$ is the gravitational constant. The equations of state (EOS)
\begin{equation}
\rho_i = f_i\left(P, T\right)
\end{equation}

\noindent relates the density $\rho\left(m\right)$ to the pressure $P\left(m\right)$ and temperature $T\left(m\right)$ within each distinct chemical layer $i$. We integrate Equations~\ref{eqn:dr} and~\ref{eqn:dP} imposing that both $P$ and $r$ are continuous across layer boundaries. The outer boundary conditions on the pressure and optical depth are calculated  following the procedure described in \citet{Rogers&Seager2010ApJ} so as to take into account the `transit radius effect' \citep{BaraffeEt2003A&A}.  We then solve iteratively for the core-mantle mass ratio that yields a consistent solution (for a given distribution of mass in the outer layers, set of atmospheric parameters, total planet mass and radius). This is the common general approach for modeling planet interiors \citep[e.g.][]{ValenciaEt2006Icarus, FortneyEt2007ApJ, AdamsEt2008ApJ, Zeng&Seager2008PASP, BaraffeEt2008A&A, FigueiraEt2009A&A, GrassetEt2009ApJ}. The model we use in this work is improved over \citet{Rogers&Seager2010ApJ} by including a temperature-dependent water EOS and is different from most previous planet interior models by providing quantitative constraints on the range of plausible gas envelope masses for a given planet mass and radius. 

The thermal profile of our model planets is divided into three regimes: an outer radiative regime in the gas/fluid envelope, an inner convective regime in the gas/fluid envelope, and a solid interior in which thermal effects are neglected. We assume that in the outermost region the planets' gas/fluid envelopes are in radiative equilibrium, and use a temperature profile derived from an analytic ``two-stream" solution to the gray equations of radiative transfer for a plane-parallel irradiated atmosphere \citep[Equation (45) in][]{Hansen2008ApJS}. The \citet{Hansen2008ApJS} temperature profile describes the temperature $T\left(\tau\right)$ as a function of the optical depth, $\tau$, and depends upon the degree of stellar insolation, internal luminosity of the planet, and the ratio of the thermal to visible opacities (parameterized by $T_0$, $T_{\rm{eff}}$, and $\gamma$, respectively). The onset of convective instabilities $\left(0<\left({\partial \rho}/{\partial s}\right)_P{ds}/{dm}\right)$ delimits the transition to the convective layer of the fluid envelope. In the convective regime, we adopt an adiabatic temperature profile. Finally, within the solid ice, mantle, and core of the planets we neglect the temperature dependence of the EOSs, employing an isothermal temperature profile.  At the high pressures found in the solid interior layers, thermal corrections have only a small effect on the mass-density \citep{SeagerEt2007ApJ}.

We choose both the fiducial values and uncertainty ranges for the atmospheric parameters in our model ($T_0$, $T_{\rm{eff}}$, and $\gamma$) following  the prescription described in \citet{Rogers&Seager2010ApJ}. As our fiducial value of $T_0$, we take 558~K, the equilibrium temperature of GJ~1214b assuming full redistribution and neglecting reflection. We also consider a range of $T_0$ values from 789 to 501~K, reflecting uncertainties in the planet's albedo and in the degree to which energy is redistributed within the planet's gas envelope. Planetary bond albedo values up to 0.35 are considered. The parameter $T_{\rm{eff}}$ describes the intrinsic luminosity of the planet. GJ~1214's old-disk kinematics and lack of chromospheric activity suggest that it has a stellar age between 3 and 10~Gyr \citep{CharbonneauEt2009Nature}. We employ a simple approximate scaling relation, derived from a power-law fit to cooling calculations from \citet{BaraffeEt2008A&A}, to relate the planetÕs intrinsic luminosity to its mass, radius, and age. For a planetary age of 3-10~Gyr, we estimate a plausible range for the intrinsic luminosity of the planet $T_{\rm{eff}} = 44-69~\rm{K}$, while adopting a fiducial value of $T_{\rm{eff}}=61~\rm{K}$ corresponding to 4.5~Gyr. For gas envelopes composed of some mixture of H and He, we adopt a fiducial value of $\gamma=1$ and consider a range from $\gamma=0.1$ to 10. Because the opacity of water is far higher in thermal wavelengths than in the visible, we expect that $\gamma<1$ in a water envelope. For our water vapor atmospheres we thus adopt a fiducial value $\gamma=0.1$ and consider an uncertainty range from $\gamma=0.01$ to 1.

In the solid (uniform temperature) layers of the planet we employ EOS data sets from \citet{SeagerEt2007ApJ} which were derived by combining experimental data at $P\lesssim200~\rm{GPa}$ with the theoretical Thomas-Fermi-Dirac equation of state at high pressures, $P\gtrsim10^4~\rm{GPa}$. We consider Fe $\left(\epsilon\right)$ \citep{AndersonEt2001GeoRL}, $\rm{Mg}_{1-\chi}\rm{Fe}_{\chi}\rm{SiO}_3$ perovskite \citep{ElkinsTanton2008E&PSL}, and H$_2$O ice \citep{HemleyEt1987Nature}. To describe hydrogen and helium  envelopes we use EOSs for H/He mixtures from \citet{SaumonEt1995ApJS} and opacity tables from \citet{FreedmanEt2008ApJS} and \citet{FergusonEt2005ApJ}. We have compiled a temperature-dependent EOS for water up to 32 GPa spanning liquid, vapor, super-fluid, and plasma phases. Our water EOS combines data from the ``The IAPWS Formulation 1995 for the Thermodynamic Properties of Ordinary Water Substance for General and Scientific Use" \citep[IAPWS-95;][]{Wagner&Pruss2002JPCRD} retrieved from the NIST Chemistry WebBook \citep{NISTwebbook},  extrapolations of the IAPWS-95 formulation calculated using FLUIDCAL software \citep{FLUIDCAL}, and the Thomas-Fermi-Dirac EOS~\citep{Salpeter&Zapolsky1967PhysRev}. For the opacity in the water vapor layer we use Planck means calculated with molecular line data from \citet{FreedmanEt2008ApJS}.  

\section{Results}
\label{sec:results}

\subsection{Background}

We consider three cases for the interior makeup of GJ~1214b: I) a planet that formed from nebular gas, ice, and rock and still harbors a primordial H/He envelope; II) an ice-rock planet that failed to accrete H/He gas from the protoplanetary disk but now has a vapor envelope; III) a  rocky planet  with an outgassed atmosphere (Figures~\ref{fig:flowchart} and \ref{fig:prism}). These scenarios determine what distinct chemical layers we consider in our differentiated planet model described in Section~\ref{sec:model}. We assume $Y=0.28$ H/He for the nebular gas, and pure H$_2$O ice for the icy material. We model the rocky material  by a combination of metallic iron and $\rm{Mg_{0.9}Fe_{0.1}SiO_3}$ silicates without imposing any a priori constraints on the iron-to-silicates ratio. We assume that, during GJ~1214b's formation, the primordial rocky material differentiated to form an iron core and silicate mantle in the planet. In this way, the rocky material contributes two layers in our planet interior structure model. The effect of choosing other chemical compositions to represent the primordial gas, ice, and rock is discussed in Section~\ref{sec:mu}.
 
We say that an interior composition is consistent with the measured planetary mass and radius within their $n\sigma$ observational uncertainties if there is some choice of planet mass within $\left(M_p-n\sigma_{M_p}, M_p+n\sigma_{M_p}\right)$, planet radius within $\left(R_p-n\sigma_{R_p}, R_p+n\sigma_{R_p}\right)$, and atmospheric parameters ($T_{\rm{eff}}$, $T_0$, and $\gamma$) within the ranges given in Section~\ref{sec:model} that yields a consistent solution for the interior composition. A more sophisticated error analysis \citep[such as that described in][]{Rogers&Seager2010ApJ} is not yet warranted given the current error bars on GJ~1214b's mass and radius. 
 
We employ ternary diagrams to graphically present our composition constraints for each scenario (Figures~\ref{fig:neptunetern} and \ref{fig:wptern}). Ternary diagrams are useful tools to graphically represent three component data $\left(x,y,z\right)$ for which the components are constrained to be positive $\left(x,y,z\geq0\right)$  and to have a constant sum $\left(x+y+z=1\right)$. Since such data have only 2 degrees of freedom, it could easily be displayed with a x-y Cartesian plot, wherein the axes and the $z=0$ line would form a right triangle. To show all three components $\left(x,y,z\right)$ on an equal footing, the x-y Cartesian plot can be squished (via a linear transformation) so that the x and y axes meet at a $60^{\circ}$ angle and form an equilateral triangle with the $z=0$ line. The resulting equilateral triangle diagram is a ternary diagram. The three vertices of the diagram represent points where $x=1$, $y=1$, and $z=1$, while $x=0$, $y=0$, and $z=0$ along the respective opposing edge. At each interior point, the value of $x$ is given by the perpendicular distance from the $x=0$ edge, with the values of $y$ and $z$ defined analogously. More detailed descriptions of how to read ternary diagrams can be found in \citet{ValenciaEt2007bApJ} and \citet{Zeng&Seager2008PASP}. 
	
\subsection{Case I: Gas-Ice-Rock Planet with Primordial Gas Envelope}

We first consider the case in which GJ~1214b managed to acquire and retain H/He gas from the protoplanetary disk. In this scenario GJ~1214b incorporated primordial iron, silicates, ice, and gas into its bulk make-up. We thus allow for four chemically distinct layers in the planet interior: an iron core, silicate mantle, water-ice layer, and H/He envelope (with $Y=0.28$).

\begin{figure}
\plotone{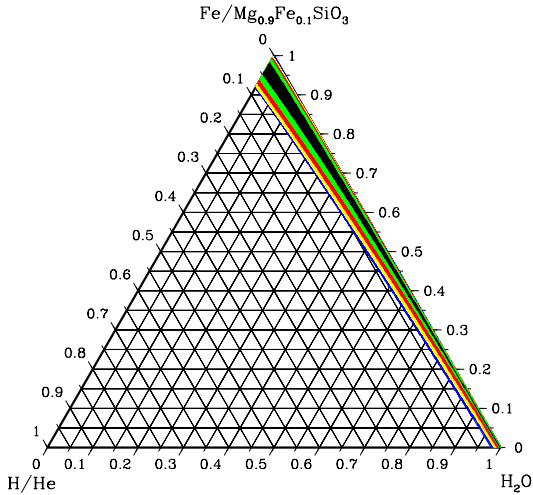}
\caption{Ternary diagram for case I in which GJ~1214b retained a primordial H/He envelope, having formed from primordial gas, ice, and rock. The relative contributions of the iron core,  $\rm{Mg_{0.9}Fe_{0.1}SiO_3}$ mantle, $\rm{H_2O}$ ices, and H/He envelope to the mass of the planet are plotted. The core and mantle are combined together on a single axis, with the vertical distance from the upper vertex determined by the fraction of the planet's mass in the two innermost planet layers. The black shaded region denotes the interior compositions that are consistent with the nominal planet mass and radius ($M_p=6.55~M_{\oplus}$,  $R_p=2.678~R_{\oplus}$) for our fiducial choice of atmospheric parameters ($\gamma=1$, $T_0=558~\rm{K}$, $T_{\rm{eff}}=61~\rm{K}$). The H/He mass fraction has a spread in this case due to the range of possible core-to-mantle mass ratios. The span of plausible interior compositions widens to the green shaded area when the range of atmospheric parameter values delimited in Section~\ref{sec:model} is considered. The red, yellow, and blue shaded regions denote compositions that are consistent with $M_p$ and $R_p$ to within 1, 2, and $3\sigma$  of their observational uncertainties, respectively, when uncertainties in the atmospheric parameters are also included.}
\label{fig:neptunetern}
\end{figure}

 Despite the range of allowed compositions, the mass of GJ~1214b's H/He gas envelope is tightly constrained (Figure~\ref{fig:neptunetern}). A gas mass fraction of 0\% is not allowed within the $1\sigma$ observational error bars on $M_p$ and $R_p$. The maximum and minimum gas envelope masses in this scenario occur for end-member mass distributions where the gas envelope surrounds a pure iron interior and a pure water ice interior, respectively. For  $M_p=6.55~M_{\oplus}$,  $R_p=2.678~R_{\oplus}$, the mass in the gas layer could be at most 3.6\%-5.2\% of the planet mass depending on the atmospheric thermal profile (hotter atmospheres require less mass of H and He to occupy a similar volume). When $M_p$ and $R_p$ are varied within their $1\sigma$ observational uncertainties, the range of maximum H/He mass fractions widens to 3.2\%-6.8\%. The minimum gas mass fraction is more sensitive than the maximum to  the atmospheric energy budget and composition (e.g., metallicity and opacities) and could be in the range $9\times10^{-5}$ and $2\times10^{-3}$ at the fiducial planetary mass and radius. 

In this scenario, GJ~1214b has  a less massive H/He envelope than our solar system Neptune, whose composition is roughly  5\%-15\% H and He, 60\%-70\% ices, and 25\% rocks by mass \citep{PodolakEt1991book, HubbardEt1995conf}. GJ~1214b could, nonetheless, support a gas envelope that is large as compared to the terrestrial solar system planets. For a Ganymede-like interior with iron:silicates:water ice in the ratio 3:22:75 by mass, GJ~1214b requires an H/He envelope accounting for between 0.01\% and 0.6\% of the planetary mass depending on the atmospheric temperature. This is up to 60 times larger than the atmosphere mass fraction on Venus ($\sim10^{-4}$).

\subsection{Case II: Ice-Rock planet with Sublimated Vapor Envelope}

A planet interior dominated by ice and a concomitant  gas envelope dominated by vapors from ice-forming materials is an intriguing possibility for  GJ~1214b. This scenario is substantially different from any of the solar system planets, but could be thought of as a class of bigger, hotter versions of Jupiter's icy moons. \citet{Kuchner2003ApJ} and \citet{LegerEt2004Icarus} first proposed that water-rich planets might be prevalent on orbits accessible to transit and radial velocity detections, although no such planets have been conclusively discovered so far. The formation pathway for these planets involves inward migration of proto-planets that formed from volatile ice-rich material beyond the snow line but that never attained masses sufficient to accrete large amounts of H/He nebular gas. If GJ~1214b falls into this category, it could be the first discovered member of a whole new population of exoplanets.

For this scenario, in which GJ~1214b did not accrete or retain any H or He from the protoplanetary disk,  we adopt an interior planet structure consisting of an iron core, silicate mantle, and water envelope. The pressure-temperature (PT) profile determines the phase of water in the envelope; vapor, liquid, super-fluid, high pressure ices, and plasma phases are all included in our H$_2$O EOS. We model the thermal profile of the water envelope following the same prescription as for the H/He layers (described in Section~\ref{sec:model}). To allow for the presence of a greenhouse effect, we take  $\gamma=0.01-1$ for H$_2$O  (compared to $\gamma=0.1-1.0$ for H and He).  

\begin{figure}
\plotone{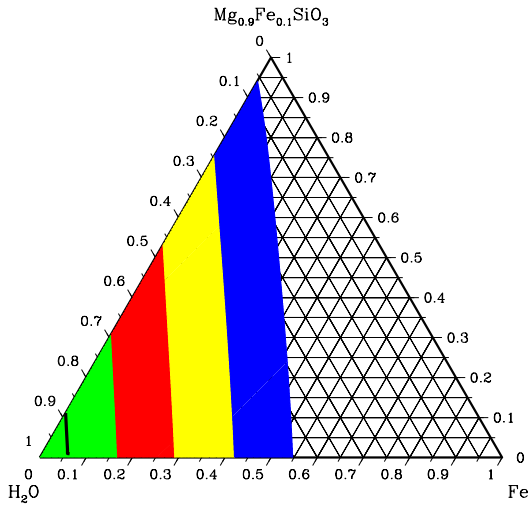}
\caption{Ternary diagram for case II in which GJ~1214b formed from refractory material and ices and has an envelope dominated by vapor from sublimated ice. The fractions of the planet's mass in the Fe core, the $\rm{Mg}_{0.9}\rm{Fe}_{0.1}\rm{SiO}_3$ silicate mantle, and the water vapor envelope are plotted on the three axes. The solid black curve represents the locus of interior compositions that are consistent with the nominal planetary mass and radius ($M_p=6.55~M_{\oplus}$,  $R_p=2.678~R_{\oplus}$) for our fiducial choice of atmospheric parameters ($\gamma=1$, $T_0=558~\rm{K}$, $T_{\rm{eff}}=61~\rm{K}$). The colors in this figure have the same designations as in Figure~\ref{fig:neptunetern}.}
\label{fig:wptern}
\end{figure}

We show in Figure~\ref{fig:wptern} the possible distributions of mass between the core, mantle, and water envelope that are consistent with the measured mass and radius of GJ~1214b. A sublimated vapor dominated envelope on GJ~1214b is possible if water accounts for a large fraction of planet mass. At the fiducial measured planet mass and radius ($M_p=6.55~M_{\oplus}$,  $R_p=2.678~R_{\oplus}$) at least 88\% H$_2$O by mass is required. To account for the observed planet mass and radius within their 1, 2, and $3\sigma$ observational uncertainties, at least 47\%, 24\%, and 6\% water by mass are required, respectively. \citet{LegerEt2004Icarus} and \citet{SelsisEt2007Icarus} have put forward that ice-rock planets formed beyond the snow line may have a comet-like bulk composition with 50\% H$_2$O and 50\% silicates and iron by mass. This composition is consistent with the measured GJ~1214b radius within $1\sigma$. Interior structure considerations do not preclude the possibility that GJ~1214b is water rich.

GJ~1214b does not contain liquid water in any of our model interiors displayed in Figure~\ref{fig:wptern}. The PT profiles that result from  the range of equilibrium temperature and internal heat flux ($T_0$ and $T_{\rm{eff}}$) values we considered are too hot to allow liquid water, even at high pressures. Our putative GJ~1214b water envelopes begin in the vapor phase at low pressures, then continuously transition to a super-fluid at  $P=22.1~\rm{MPa}$ (the critical pressure of water), before eventually becoming an electronically conductive dense fluid plasma at greater depths ($T\gtrsim4000~\rm{K}$).

To obtain liquid water in our model interior, we must decrease GJ~1214b's assumed intrinsic energy flux  below  $5\times10^{-4}~\rm{W~m^{-2}}$ or  $T_{\rm{eff}}\lesssim 10~\rm{K}$; for comparison, Earth's internal heat flux is $0.087~\rm{W~m^{-2}}$ \citep{Turcotte&Schubert2002book}. The intrinsic luminosity of GJ~1214b is very uncertain, and a detailed evolution calculation to better constrain its magnitude is out of the scope of this work. Nonetheless, we predict that, in this scenario, GJ~1214b would need a cold interior in order to harbor liquid water.

Our conclusions regarding the possibility of liquid water in an extended vapor atmosphere on GJ~1214b are contingent upon our parameterized PT profile adequately describing the water envelope. It is important to note that our model thermal profile assumes GJ~1214b is in radiative equilibrium with the incident stellar irradiation it receives at its current orbital location. If GJ~1214b has recently migrated and is undergoing active vaporization (as in the scenarios considered by \citet{Kuchner2003ApJ}  and \citet{ValenciaEt2009AstroPh}) it could have temperature and H$_2$O-phase profiles within its envelope that are very different from those available within the framework of our models (e.g., atmospheres with a temperature inversion or those out of radiative-convective equilibrium).

\subsection{Case III: Rocky Planet with Outgassed Atmosphere}
\label{sec:case3}

We turn to  the possibility that GJ~1214b formed from purely rocky material without retaining any H/He gas or icy material from the protoplanetary disk. We reiterate that GJ~1214b must still have a substantial gas layer in this case, because a rocky planet is too dense to account for the measured mass and radius (e.g., a gasless Earth-like composition yields a planet radius that is $8\sigma$ lower than that measured for GJ~1214b). In this rocky planet scenario, an outgassed atmosphere contributes to the GJ~1214b transit radius. We focus here on an atmosphere produced by outgassing during planet formation; outgassing from post-formation geological activity is another possibility, which is discussed in Section~\ref{sec:oo}.

It is difficult to predict a priori the composition of the gas layer that would be produced by outgassing on GJ~1214b. The initial composition of the outgassed atmosphere is strongly dependent on the composition of planetesimals comprising GJ~1214b. Even among solar system chondrites, the outgassed atmosphere compositions would range from H$_2$-dominated, to H$_2$O-dominated, to CO-dominated, to CO$_2$-dominated \citep{Schaefer&Fegley2009bAstroPh}. In addition, the conditions during magma solidification and outgassing affect the gas layer outcome \citep{ElkinsTanton&Seager2008aApJ, Schaefer&Fegley2009bAstroPh}. Further complicating the rocky planet gas layer, is the subsequent atmospheric escape \citep{CharbonneauEt2009Nature} and photochemistry. Despite this uncertainty in composition, we can nevertheless make some concrete statements about a putative outgassed envelope on GJ~1214b. 

If GJ~1214b's gas layer was produced by outgassing, a substantial fraction of it must be in a component that is less dense than water vapor. The light component is needed for GJ~1214b's envelope to have both a mass low enough to be produced by outgassing and a volume large enough to account for the transit radius. From our study of case II, we found that if water is the least dense component of GJ~1214b, at least 47\% H$_2$O by mass is required within $1\sigma$ (Figure~\ref{fig:wptern}). This is more water than a terrestrial planet formed from chondritic planetesimals can degas (up to 23\% by mass) \citep{ElkinsTanton&Seager2008aApJ}. Volatile molecules heavier than water (such as CO, CO$_2$, and N$_2$) have even smaller upper bounds (in \% planet mass) on the amounts they can be outgassed. The requirement for substantial quantities of a light species is interesting because it limits the outgassed atmosphere compositions that can be relevant to GJ~1214b from among the wide range of a priori possibilities. In particular, an Earth-like N$_2$-dominated outgassed atmospheric composition and a Venus-like CO$_2$ outgassed atmosphere are both ruled out for GJ~1214b. 

Molecular hydrogen, H$_2$, is the most likely candidate for an atmospheric species that is both light enough and  plausibly outgassed in sufficient quantities to account for GJ~1214b's transit radius. In fact, H$_2$ is predicted to dominate the atmospheres outgassed by ordinary H, L, LL, and high iron enstatite EH chondrite-composition planetesimals \citep{ElkinsTanton&Seager2008aApJ, Schaefer&Fegley2009bAstroPh}.  Although He has a low molecular weight, it does not bind to minerals the way H does, and consequently it cannot be accreted with the rocky primordial material and later released through outgassing \citep{ElkinsTanton&Seager2008aApJ}. The other possible species, CH$_4$ and NH$_3$, have molecular weights only slightly lower than water and are not typically outgassed in large quantities.

\begin{figure}
\plotone{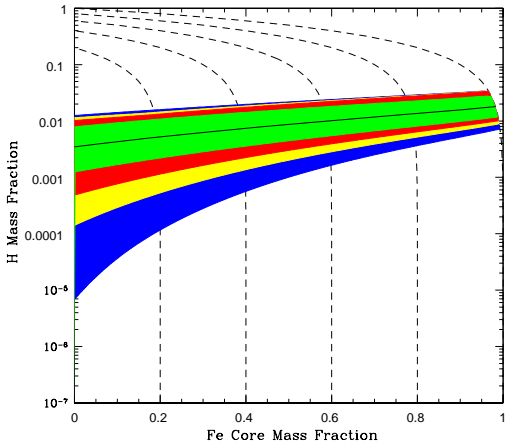}
\caption{GJ~1214b as a rocky planet with an outgassed atmosphere. A three-layered planet structure composed of an iron core, silicate mantle, and pure hydrogen envelope is assumed. The fraction of the planet's mass in the hydrogen gas envelope is plotted as a function of the fraction of the planet's mass in the iron core. Any mass not in the hydrogen envelope or iron core is contained in the silicate mantle; the dashed black lines represent contours of constant mantle mass fraction. The colored shaded regions have the same designations as in Figure~\ref{fig:neptunetern}. }
\label{fig:hatmplot}
\end{figure}

We emphasize that only a relatively small ``out-gassable" amount of H$_2$ is required to account for GJ~1214b's transit radius. We show this quantitatively by considering a pure hydrogen gas layer surrounding a rocky interior in Figure~\ref{fig:hatmplot}.  Just $5\times10^{-4}$ of the planet mass in H$_2$ surrounding a rocky core is sufficient to account for the transit radius to within $1\sigma$. This is 2 orders of magnitude below the maximal amount of hydrogen that might be outgassed by a rocky planet (6\% of the planet mass), which corresponds to the extreme where the planet formed from planetesimals similar to EH chondrites and fully oxidized during formation \citep{ElkinsTanton&Seager2008aApJ}. For an iron core mass fraction similar to Earth's (30\%), GJ~1214b needs 0.3\%-1.2\% of its mass in a pure H$_2$ gas layer at the nominal mass and radius. This H$_2$ gas layer is small compared to the H$_2$O envelopes in the water planet scenario, but is still large compared to Earth's atmosphere (which accounts for roughly 0.0001\% of Earth's mass).

A pure H$_2$ envelope as we have assumed above is somewhat artificial; realistically H$_2$ will not be outgassed on its own but in combination with heavier molecules. At a given Fe core mass fraction, including H$_2$O, CO, CO$_2$, and other additional species in the outgassed envelope would in general tend to decrease the mass fraction of H$_2$ required to reproduce the planet radius (Figure~\ref{fig:hatmplot}), while increasing the total gas mass fraction of all volatile species combined. This is because {\it i)} the heavier outgassed volatiles are still less dense than silicates and {\it ii)} the additional atmospheric species would increase the opacity of the atmosphere. This strengthens the result that an outgassed hydrogen-rich envelope surrounding a rocky solid interior is not ruled out by the measured mass and radius of GJ~1214b. We consider how this interpretation is influenced when atmospheric escape is taken into account in Section~\ref{sec:ae}.

\section{Discussion}
\label{sec:dis}

\subsection{Atmosphere Observations}

What further observations can discriminate among the gas layer origin possibilities? A promising possibility is spectral observations of the planetary atmosphere. Here, we summarize some ideas for how atmospheric spectra might help to distinguish between a gas envelope dominated by nebular gas, sublimated ices, or outgassing.

 Atmospheric He is the discriminator between captured H/He envelopes or outgassed envelopes \citep{ElkinsTanton&Seager2008aApJ}. He is difficult to observe spectroscopically at planetary temperatures, however. Recent outgassing by geological activity could be revealed through the detection of spectral features from species with short photochemical lifetimes such as SO$_2$ \citep{KiteEt2009ApJ}. A water vapor dominated atmosphere would have saturated water vapor features (e.g., absorption bands at $\lambda\sim5-8\mu\rm{m}$ and $\lambda\sim16-18\mu\rm{m}$ \citet{LegerEt2004Icarus}), but these may not be unique identifiers.  Discriminating between a hydrogen-rich envelope and a water vapor atmosphere on the basis of the atmospheric scale height \citep{MillerRicciEt2009ApJ} might be more promising than discriminating on the basis of the presence or absence of spectral lines. The scale height is accessible via the depth of strong (or saturated) transit transmission spectral features. Although it will be tricky to constrain interior compositions from atmospheric spectra, the prospect of greater insights into the composition and formation of GJ~1214b and its envelope is compelling. 
 
 \subsection{Atmospheric Escape}
 \label{sec:ae}
 
We cannot definitely rule out any of our three composition scenarios on the basis of atmospheric escape calculations. The water vapor envelope in case II is most resilient against atmospheric escape. While mass loss is a more important consideration for the hydrogen-rich primordial and outgassed envelopes in cases I and III, uncertainties in the mass-loss rate and planet age leave room for the possibility that GJ~1214b retained sufficient H/He or H$_2$ to account for its transit depth. Further observations are warranted to tighten the constraints on the origin and character of GJ~1214b's gas layer.
 
 Atmospheric escape may have an important influence on mass and composition evolution of the GJ~1214b gas envelope.  We calculate $2\times10^6~\rm{kg\,s^{-1}}$ for an energy-limited upper bound on the current mass-loss rate of GJ~1214b following the approximate approach of \citet{LeCavelierDesEtangs2007A&A}. This is on the same order as the  $9\times10^5~\rm{kg\,s^{-1}}$ escape rate \citet{CharbonneauEt2009Nature} predicted for a hydrogen-rich atmosphere escaping hydrodynamically from GJ~1214b. At  $2\times10^6~\rm{kg\,s^{-1}}$, $0.03~M_{\oplus}$ would be lost over 3~Gyr and $0.1~M_{\oplus}$ would be lost over 10~Gyr.  The actual cumulative mass lost over GJ~1214b's lifetime may be much higher than these values because the host star was probably brighter in UV at earlier times.
 
Out of the three cases we considered for GJ~1214b's composition, case II, the ice-rock scenario having a water vapor envelope, is most robust against atmospheric escape. First, water has a higher molecular weight than H or He which makes it easier to retain (although photodissociation of H$_2$O may be important). Second, with H$_2$O comprising upward of 50\% of GJ~1214b's current mass in this scenario, a cumulative water loss on the order of $0.1~M_{\oplus}$ will not have significantly changed the overall character of the planet.  In contrast, losses of this magnitude are far more significant for the hydrogen-rich envelopes considered in cases I (primordial gas dominated) and III (outgassing dominated), since these envelopes can account for at most a few percent of GJ~1214b's current mass.

 In case I, atmospheric escape increases the amount of gas GJ~1214b would have needed to accrete from the protoplanetary nebula. Protoplanetary cores between 1 and $10~M_{\oplus}$ can acquire substantial primordial atmospheres even if the cores are too small for the nucleated instability and runaway gas accretion to commence \citep{Mizuno1980PThPh}. For instance, \citet{Rafikov2006ApJ} found that a $6.5~M_{\oplus}$ core in a minimum mass solar nebula could develop an atmosphere of up to several tenths of  an Earth mass of nebular gas (depending on the core's semimajor axis and accretion luminosity). Whether GJ~1214b could accrete and retain enough nebular gas to account for its transit radius is ambiguous.

Atmospheric mass loss creates the most tension with case III, because there is an upper bound on the cumulative amount of gas GJ~1214b can outgas over its lifetime. Based on solar system meteorites, a rocky planet can outgas at most 6\% of its mass as hydrogen \citep{ElkinsTanton&Seager2008aApJ} and less will be outgassed if the iron content of the rocky material is not fully oxidized. It is likely that GJ~1214b has lost most of any H-atmosphere outgassed during formation. However, only $\sim0.1\%$ of the planet mass in H$_2$ surrounding a rocky core could be needed to account for the transit radius to within $1\sigma$. Due to the large uncertainties on the planet age and time averaged mass-loss rate, the outgassing cannot be fully ruled out as the source of GJ~1214b's gas layer.

\subsection{Ongoing Outgassing}
\label{sec:oo}

If GJ~1214b is a rocky planet with an outgassed atmosphere as in case III, ongoing outgassing by geological activity could be contributing to its gas envelope. However, the rate of ongoing outgassing is expected to be smaller than the atmospheric escape rate.  \citet{KiteEt2009ApJ} predict that a $\sim6.5~M_{\oplus}$ super Earth experiencing plate tectonics would have rates of volcanism per unit planet mass 6 times Earth's current rate at planet age of 3~Gyr  and 0.2-0.3  times Earth's current rate at a planet age of 10~Gyr. Lower rates of volcanism are predicted for planets 3-10~Gyr old that are not tectonically active. If we consider the most optimistic case for the rate of volcanism of GJ~1214b and assume a magma volatile content similar to volatile-rich terrestrial magmas at mid-ocean ridges (1.5\% H$_2$O and $400~\rm{ppm}$ CO$_2$ by mass \citet{Oppenheimer2003TrGeo}), we obtain an upper limit on the volcanic outgassing rate of $1\times10^{6}~\rm{kg\,s^{-1}}$ H$_2$O and  $4\times10^{4}~\rm{kg\,s^{-1}}$ CO$_2$. Although outgassing by geological activity may help to replenish the GJ~1214b envelope, it probably cannot completely offset the effect of atmospheric escape if the planet is older than 3~Gyr. 

\subsection{Necessity of a Gas Layer}

GJ~1214b's low average density implies that it has a low density gas  envelope.  This statement is valid as long as the true planet mass and radius lie within the $2\sigma$ measurement uncertainties. If GJ~1214b's mass and radius both differ from their measured values by more than $\sim2\sigma$, a solid,  ice-dominated interior composition with no gas envelope is barely allowed. A pure ice planet seems physically implausible, however, because silicates (i.e., higher density material) are expected to be accreted along with ices during planet formation. \citet{CharbonneauEt2009Nature} point out, however, that the stellar radius they derived for GJ~1214 from observations is 15\% larger than that predicted by the theoretical models \citet{BaraffeEt1998A&A}. If systematics have led to an overestimation of the planet radius, the evidence for a gas layer on GJ~1214b would be reduced.   

\subsection{Model Uncertainties}
\label{sec:mu}

Despite our quantitative constraints on the range of interior compositions, we are faced with some uncertainties. One uncertainty is the atmospheric temperature, controlled by the unknown interior energy and unknown albedo. A hotter atmosphere fills a larger volume than a cooler atmosphere, requiring less atmospheric mass to fit the planet radius. We have chosen a range of reasonable values for the parameters governing the atmospheric PT profile (Section~\ref{sec:model}) and found that the model uncertainty is  roughly comparable to the observational uncertainties. Although we adopted nearly identical atmospheric parameter ranges for all three interior structure cases we considered, in reality GJ~1214b's albedo, atmospheric absorption, and interior luminosity are coupled to its interior composition.

The precise chemical make-up of GJ~1214b's interior layers is another source of uncertainty. We have adopted artificially clean single-chemical materials to represent GJ~1214b's core (pure Fe), mantle ($\rm{Mg_{0.9}Fe_{0.1}SiO_3}$ perovskite), and ice layer (pure H$_2$O). In reality, we expect a mixture of chemical compounds to contribute to each layer. The presence of a light element (such as sulfur) in the iron core would decrease the mass required in the low-density outer envelope. In contrast, a higher iron content in the mantle would increase the density of the planet interior, requiring a higher mass fraction of gas.  While H$_2$O should dominate any primordial ices forming GJ~1214b (contributing more than $\gtrsim60\%$ by mass), substantial quantities (more than 1\% by mass) of  CO, CO$_2$, H$_2$S, NH$_3$, N$_2$, and CH$_3$OH are also expected in ices formed from a protoplanetary disk of roughly solar composition \citep{MarboeufEt2008ApJ}. The presence of significant quantities of CO$_2$ could have dramatic effects on the evolution and thermal structure of a water planet, maintaining the steam atmosphere in a hot state \citep{LegerEt2004Icarus}.  It is important to note that the range of possibilities for the chemical materials comprising GJ~1214b's interior is constrained by the cosmic abundance of the elements. We have focused on possible silicate-based interior composition scenarios for GJ~1214b. Alternatively, SiC, graphite, and other carbon compounds could dominate the interiors of planets formed under conditions where $\rm{C/O}>1$ by number \citep{Kuchner&Seager2005AstroPh}, but we have not considered this possibility here.

We find that  the choice of opacities used to model the GJ~1214b envelope has a considerable effect on the H/He mass fraction constraints in case I.  \citet{FreedmanEt2008ApJS} opacities were used to generate the composition constraints shown in  Figure~\ref{fig:neptunetern}. The \citet{FergusonEt2005ApJ} opacities, which include condensate and grain opacity sources, tend to be higher overall than the \citet{FreedmanEt2008ApJS} opacities, which do not include grain opacity sources. Consequently, gas envelopes modeled with the \citet{FergusonEt2005ApJ} opacities systematically require less H/He mass to reproduce the observed transit radius than those modeled with the \citet{FreedmanEt2008ApJS} opacities. This illustrates that the relationship between the radial thickness of an H/He envelope and its mass is sensitive to the precise metallicity and composition assumed. Despite the added uncertainty introduced by the metallicity of the H/He layer  our inferences regarding the plausibility of a primordial gas envelope on GJ~1214b remain unchanged.

\section{Summary and Conclusions}
\label{sec:con}

The MEarth transiting planet, GJ~1214b, is exciting because it lies in a mass and density regime for which there are no solar system analogs; GJ~1214b is smaller than the ice giants Neptune and Uranus, while larger than the terrestrial Earth, Venus, and Mars. We emphasize that, based on its measured planetary mass and radius alone, we can constrain GJ~1214b's composition but we cannot infer its unique true composition. 

GJ~1214b requires a gas envelope to account for its low average density so long as the true planet mass and radius lie within $2\sigma$ of their measured values. With interior structure models, we explored three possible scenarios for the gas layer and concomitant  interior of GJ~1214b. An important conclusion from this investigation is that, under most of the conditions we considered, GJ~1214b would not have liquid water. We summarize more detailed results for each of the three cases below. 

If GJ~1214b's gas layer was accreted directly from the protoplanetary nebula, the primordial H/He layer surrounding an interior of iron, silicates, and ice would need to contain between 0.01\% and 5\% of the planet mass in order to account for the transit radius. This is interesting because the gas envelope would be less massive than Uranus' and Neptune's envelopes (which account for 5\%-15\% of the planet mass), yet greater than Earth's or Venus' atmospheres (which contribute $0.0001\%$ and $0.01\%$ of the planet mass, respectively). 

If, instead, sublimated ices dominate the gas layer, a massive water envelope comprising at least 47\% of the planet mass could account for GJ~1214b's observed parameters to within $1~\sigma$. We thus do not require an H/He layer to explain the measured mass and radius. In this sublimated ice-dominated case, for our assumptions about the planet albedo ($A\leq0.35$) and internal heat flux, we find that GJ~1214b's water envelope would generically be too hot to allow liquid water, even at high pressures.  Instead of a liquid water layer the planet would have a super-fluid water layer sandwiched between plasma below and vapor above. To obtain liquid water in our model interior, we must decrease GJ~1214b's assumed intrinsic energy flux to $\lesssim0.01$ of Earth's intrinsic energy flux. 

Third, if the nascent GJ~1214b did not manage to retain any primordial gas or ices, outgassing from rocky material could produce a gas layer surrounding a terrestrial interior. In order for sufficient gas to be released in this scenario to account for the transit radius, the outgassed atmosphere would need to be hydrogen rich. This in turn constrains the mineralogy of the primordial rocky material from which GJ~1214b would need to have formed in this case. Based on models of outgassing of chondritic material found in the solar system, even if hydrogen is the dominant outgassed species other heavier volatiles would be present as well. It is also expected that atmospheric escape would have eroded a substantial amount of the H atmosphere \citep{CharbonneauEt2009Nature}, which provides added tension with this scenario. Although outgassing is not ruled out as the primary contributor of GJ~1214b's gas layer, more modeling is required to verify its viability.

The ideal hope is that we may gain insights into planet migration and formation if future observations succeed in constraining the composition of GJ~1214b's atmosphere and then by extension the planet's interior. Despite the link between atmospheres and interiors, we caution that the solar system planet atmospheres, especially terrestrial atmospheres that have undergone substantial evolution, are divorced from their interiors. Progress toward constraining GJ~1214b's interior might thus be challenging. Additional complications arise from the fact that multiple processes may have contributed to the gas layer on GJ~1214b. Nevertheless, here are some hopeful possible outcomes.  If GJ~1214b's gas layer is found to have close to solar abundance of He, then  it is likely composed of nebular gases, and together with atmospheric escape estimates may yield bounds on the accretion of gas from protoplanetary disks by small planetesimals. If instead sublimated ices dominate with no observable H and He, it would indicate that GJ~1214b formed beyond its star's snow line and migrated inward to its current orbital distance. Finally, if GJ~1214b's envelope has no He but a substantial amount of H and other volatiles, it would likely be the result of outgassing, and may help to constrain its atmospheric mass-loss rate and the broad-brush oxidation properties of the rocky material from which GJ~1214b formed. No matter which atmospheric sources turn out to be dominant on GJ~1214b (be it nebular gas, sublimated vapors, or outgassing) we will learn something interesting, and perhaps achieve our first glimpse of a planetary atmosphere on a whole new class of low-mass exoplanets.

\acknowledgments

We thank Lisa Messeri and Josh Winn for very helpful discussions. We thank Marc Kuchner for his referee comments, which were very insightful and helped to strengthen this article. We also thank Richard Freedman for providing molecular line lists. This work is supported by the Natural Sciences and Engineering Research Council of Canada.

\bibliography{exoplanets}

\clearpage

\end{document}